\documentclass[11pt]{article}
\begin{document}

\begin{center}
{\bf \large{Ferromagnetic properties of charged vector boson condensate
}} \\ \vspace{0.5cm}
{\it 
Alexander D. Dolgov \footnote{dolgov@fe.infn.it}$^{a,b,c}$, 
Angela Lepidi \footnote{lepidi@fe.infn.it}$^{a,b}$,
and Gabriella Piccinelli \footnote{gabriela@astroscu.unam.mx}$^{d}$
}
\\\vspace{0.3cm}
$^a$ Istituto Nazionale di Fisica Nucleare, Sezione di Ferrara, 
I-44100 Ferrara, Italy \\
$^b$ Dipartimento di Fisica, Universit\`a degli Studi di Ferrara, 
I-44100 Ferrara, Italy \\
$^c$ Institute of Theoretical and Experimental Physics, 113259 Moscow, Russia \\
$^d$ Centro Tecnol\'ogico, FES Arag\'on, Universidad Nacional Aut\'onoma de  M\'exico, 
Avenida Rancho Seco S/N, Bosques de Arag\'on, Nezahualc\'oyotl, 
Estado de M\'exico 57130,  M\'exico

\begin{abstract}
Bose-Einstein condensation of W bosons in the early universe is studied. 
It is  shown that, in the broken phase of the standard electroweak theory, 
the condensed W bosons form a ferromagnetic state with aligned spins. 
In this case the primeval plasma may be spontaneously magnetized
inside macroscopically large domains and form magnetic fields which 
may be the seeds for the observed today galactic and intergalactic fields.
However, in a modified theory, e.g. in a theory with stronger quartic self 
interactions of gauge bosons e.g. due to a smaller value of the weak 
mixing angle, antiferromagnetic condensation is possible.  In the latter 
case W bosons form scalar condensate with macroscopically large 
electric charge density i.e. with a large average value of the bilinear 
product of W-vector fields but with microscopically small average
value of the field itself. \end{abstract}
\end{center} 

\section{Introduction \label{s-intro}}

Bose-Einstein condensation is the quantum phenomenon of 
the accumulation of identical bosons
in the same state, which is their lowest energy (zero momentum) state. Under 
these conditions they behave as a single macroscopic entity described by a 
coherent wave function rather than a collection of separate independent particles. 
Even though the Bose-Einstein condensation  had been foreseen long time 
ago (1925), it took seventy years to make the first experimental observation, which was 
performed in a dilute gas of rubidium~\cite{Anderson:1995gf}. 
Difficulties in performing this observation were created by the extremal conditions  
necessary for the condensation. Indeed,
the Bose-Einstein condensation takes place when the inter-particle separation is 
smaller than their de Broglie wavelength, $\lambda_{dB} \sim 2\pi/\sqrt{2m T}$, so the 
system must be cooled down to a very low temperature at ordinary densities. 

In the recent years the study of Bose-Einstein condensate (BEC) became an active area
of research in different fields of physics from plasma and 
statistical physics (for a review see book \cite{Pethick_book} and 
references therein) to astrophysics and cosmology.

The presence of charged Bose-Einstein condensates has interesting consequences in gauge 
field theories. For instance, in our 
recent papers \cite{Dolgov:2008pe,Dolgov:2009yt}, electrodynamics of charged fermions
and scalar bosons was considered with the fermion asymmetry above a 
certain critical value such that the boson chemical potential had to be equal to the boson 
mass. In these conditions the bosons should condense to ensure
electric neutrality of the plasma. The screening of impurities in such plasma is 
essentially different from the case when the condensate is absent.

A similar problem has been recently considered also in the 
framework of an effective field 
theory \cite{Gabadadze:2008mx,Gabadadze:2008pj,Gabadadze:2009jb}. 
The authors focused on the applications to the astrophysics
of the helium white dwarfs 
discussing possible condensation of helium nuclei
and analyzing the thermodynamical properties of the system and possible 
observational signatures. The formalism they used is complementary to ours, 
but our results agree in overlaping areas.

In this work we consider the BEC of charged $W$-bosons, which may be formed  
in the early universe, and study its magnetic properties.
In general $W$ bosons may condense both
below and above the EW symmetry 
breaking if the cosmological lepton asymmetry happened to be sufficiently high, 
i.e. if the chemical potential of neutrinos was larger than the $W$ boson mass
at this temperature. The condensation of vector bosons differs from the 
theoretically simpler case of scalar bosons due to the presence of an additional 
degree of freedom, their spin states.
In both cases, scalar and vector, the condensed bosons are in the zero momentum
state but in the latter case the spins of the individual vector bosons can be either aligned or
anti-aligned. These states are called respectively ferromagnetic and 
anti-ferromagnetic ones, see e.g. ref.~\cite{Pethick_book}. 
The realization of one or the other state
is determined by the spin-spin interaction between the bosons. 
In the lowest angular momentum state, $l=0$, a pair of bosons may have either 
spin 0 or 2. Depending on the sign of the spin-spin coupling, one of those states 
would be energetically more favorable and would be realized at the condensation.
In the case of the energetically favorable higher spin state,
$S=2$, the vector bosons condense with 
macroscopically large value of their vector wave function $\langle W_j \rangle$.
In the opposite case of the favorable $S=0$ state the vector bosons
form the scalar condensate 
with pairs of vector bosons making a scalar ``particle''.  
Such phenomena are observed in solid state physics with such
spin-1 bosons as $^{23}Na$, $^{39}K$, and $^{87} Rb$ nuclei,
see refs.~\cite{Pethick_book}, \cite{Ho:1998zz}--\cite{ho-yip}. 

Usually experimental studies of
the properties of the spin-1 condensate are performed in external
magnetic fields. Under such conditions the spins of the vector bosons
are aligned (frozen) due to the interactions of their magnetic moments with the
external field. However, in optical traps an external magnetic field
is absent and spin alignment or de-alignment  depends upon the internal
dynamics of the system. Correspondingly either ferromagnetic or scalar ground 
state would be formed depending on the scattering length of vector bosons
in different angular momentum channels~\cite{Ho:1998zz}.

Let us stress again that in the ferromagnetic case the total spin of the vector bosons
has macroscopically large value, while in the antiferromagnetic case the lowest
energy state of many condensed vector bosons has zero total spin,
$S_{tot} =0$, if their number is even, and has $S_{tot} =1$ (still
macroscopically small), if their number is odd. 

Due to macroscopically large value of the total spin
in the ferromagnetic spin state the system can be
accurately described by the mean field approximation, as is argued e.g.
in sec. 12.2 of book~\cite{Pethick_book} or refs.~\cite{law-pu,ho-yip}. 
Indeed, the validity of the mean field 
approximation is determined by the relative magnitude of the fluctuations
near the ground state. The fluctuations are induced by the particle scattering
which can change the spin value in a single reaction by $\pm 1$.
It is clear that for a large value of the total spin the relative fluctuations
$\delta S/S \sim \delta N/N \sim 1/\sqrt{N} \ll 1$, while for a small total spin value 
$\delta S/S \sim 1$. 

In presence of sufficiently strong external magnetic field all the vector boson spins
condense in the same direction and thus the whole body forms a single magnetic
domain independently on the spin-spin interactions of the vector bosons, ferromagnetic
or not. If however, an external magnetic field is absent, several magnetic domains
would be formed in the ferromagnetic case, due to dynamical instability, 
and none in the antiferromagnetic case.
The discussion of this phenomenon in solid state
physics  and the list of references can be found e.g. in ref.~\cite{zhang-2005}. 

The situation  with condensation of $W$-bosons is  not as complicated
as in solid state physics where the dominant contribution
to the spin-spin interactions comes from the quantum exchange effects.
However, for the system we are considering here, 
the exchange forces are not essential and the spin-spin
interaction is determined by the interaction of the magnetic moments of the vector
bosons and by their self-interactions. 
The interactions with relativistic electrons and positrons are neglected, see below.

We show that charged vector bosons 
would condense in maximum spin states and form classical vector field, if 
only electromagnetic interactions between their spins are taken into
account. In such a case the 
spontaneous magnetization at macroscopically large scales would
take place. On the other hand, the local self-interaction of $W$
creates the spin-spin coupling of the opposite sign. In the standard
theory the magnitude of this coupling is smaller than that induced by
the photon exchange, while the exchange of heavy $Z$-boson does not
contribute at all into the spin-spin interactions of $W$-bosons. Thus 
the spin-spin coupling is dominated by the interactions of the magnetic
moments of $W$. In pure electrodynamics magnetic fields are not
screened and so one may expect that the plasma effects would not
eliminate the dominance of the  interactions between the 
magnetic moments. However,
the situation is not clear in non-Abelian theories~\cite{linde-screen} 
and in principle the screening might inhibit the
spin-spin magnetic interaction, see sec. \ref{ss-screen}.
If this is realized, the local quartic selfcoupling of $W$ would
dominate over the electromagnetic one and $W$ bosons should condense
in antiferromagnetic state and form a scalar condensate. Hence a classical
vector field would not be created.

The condensation of gauge bosons of weak interactions was considered
in the  pioneering papers~\cite{linde}, where it was argued that at
sufficiently high leptonic chemical potential a classical
$W_j$ boson field could be created in the early universe. Our results are
similar to that of ref.~\cite{linde} as long as the ferromagnetic case 
is realized. In this case the 
spins of $W$ add up coherently  creating classically large
average vector $W_j$ boson field.
In the hypothetical situation of a
stronger quartic self-coupling of $W$ 
we arrive to an opposite conclusion of vanishingly small classical 
$W$ field but with macroscopically large number density 
of $W$-bosons at rest, which is given by the bilinear product 
$n_W = i ( \partial_t W^\dagger_j W_j  - W_j^\dagger \partial_t W_j)$ 
(as we see in what follows, this expression for the
number density of $W$ is true in the gauge where the electromagnetic potential is
zero).

A different mechanism of formation of $W$-boson spin condensate by 
chaotic magnetic fields, which might exist in the early universe,  was
considered in ref.~\cite{olesen}. If such fields were sufficiently
strong, this mechanism could operate independently
on the spin-spin interaction of $W$-bosons and would align their spins
in the domains with the size of the order of that of the original
cosmic magnetic field, which are  normally microscopically small.
To this end a magnetic field with the strength $B > \alpha
m_W^2$ would be necessary. 
Such fields might be generated at the electro-weak
phase transition. The alignment of the $W$ spins reminds the
alignment of vector fields in magnetic traps mentioned above.
Moreover, such an alignment under
the influence of a sufficiently strong external magnetic field would
take place in both ferromagnetic and antiferromagnetic cases. However
when the external magnetic field is switched-off or redshifted, the spins
of the ``antiferromagnetic'' $W$-bosons would be dis-aligned making
scalar condensate, while in the ferromagnetic case  macroscopically
large domains with aligned spins would be created. The mechanism of
formation of such domains is different from the normal ferromagnets due to an
absence of the exchange forces. So probably the size of the domains
is not determined by the usual competition between the volume and surface 
energies but  by the causality effects.

A recent application of vector BEC to astrophysics 
was studied in ref.~\cite{Berezhiani:2010db}, 
where the condensation of deuterium nuclei was
considered. The authors argue that the interaction between 
deuterium nuclei forces them into the lowest spin antiferromagnetic state. 

The paper is organized as following. In sec. 2 kinetic equation approach to
BEC is considered. In sec. 3 the same phenomenon is studied on the basis of
the equations of motion of the vector fields. In sec. 4 the spin-spin interaction 
of W bosons in the zero momentum state is calculated. In sec. 5 discussion of
the results and conclusion are presented.

\section{Formation of Bose-Einstein condensate. Kinetic approach
\label{s-kinetics}}

It is convenient to describe formation of BEC using kinetic theory approach. 
Let us consider a system of bosons and fermions in thermal equilibrium with
temperature $T$. As is known, the equilibrium distribution functions,
up to the spin counting factor, are equal to:
	\begin{eqnarray}
	\label{Boson_distr}
	f_{F,B} = \frac{1}{\exp [(E-\mu_{F,B})/T] \pm 1}\,, 
	\end{eqnarray}
where the signs plus and minus stand respectively for fermions and bosons
and $\mu_{F,B}$ are chemical potentials of fermions and bosons. Evidently
the chemical potential of bosons cannot exceed their mass, $ \mu_B \leq m_B$.
Otherwise their distribution would not be positive definite. 
This upper bound on
$\mu_B$ enforces the Bose-Einstein condensation if the asymmetry between bosons and
anti-bosons is so high that even maximally large chemical potential, $\mu_B = m_B$, 
is not sufficient to provide for such a high asymmetry.
In this case the bosonic distribution function takes the form:
	\begin{eqnarray}
	\label{Bose_cond_distr}
	f_B = C \delta^{(3)} (\mathbf{p}) + \frac{1}{\exp [(E-m_B)/T] -1},
	\end{eqnarray}
where the first delta-function term describes the condensate and 
a new constant parameter $C$ is its amplitude. The second term, 
which is the usual Bose distribution, describes non-condensed particles 
and vanishes at $T=0$. 

It is easy to verify that the distributions written above annihilate the collision
integral, i.e. they are the equilibrium distributions. Indeed the kinetic equation has 
the form: 
	\begin{eqnarray}
	 \frac{df_1}{dt} = I_{coll}[f]\,,
\label{df-dt}	
\end{eqnarray}
where the collision integral is equal to
	\begin{eqnarray}
I_{coll} = -\frac{(2\pi)^4}{2E_1}\int d\tau'_{in} d\tau_{fin} 
\delta^{(4)}\left(\sum p_{in} - \sum p_{fin} \right) | A_{if} |^2 F[f],
\label{I-coll}	
\end{eqnarray}
and
	\begin{eqnarray}
	 F[f]=\Pi f_{in} (1\pm f_{fin}) - \Pi f_{fin}(1\pm f_{in})\,.
\label{F-of-f}	
\end{eqnarray}
In eq. (\ref{I-coll}) $d\tau'_{in}$ is the phase space of all initial particles 
except for particle under scrutiny, i.e. the initial particle 1, $d\tau_{fin}$ 
is the phase space
of all final particles, $A_{if}$ is the amplitude of the transition
from an initial to a final state.
The product is taken over all initial (in) and final (fin) states and signs
plus or minus stand for bosons and fermions respectively. It is straightforward to
check that $F[f]$ vanishes on distributions (\ref{Boson_distr}) for arbitrary $\mu_F$
and $\mu_B \leq m_B$ satisfying the usual condition of chemical equilibrium:
\begin{eqnarray}
\sum \mu_{in} = \sum \mu_{fin}  \,.	
\label{mu-equil}
	\end{eqnarray}
If $\mu_B = m_B$, there arises an additional freedom that $F[f]$ vanishes for 
distribution (\ref{Bose_cond_distr}) with an arbitrary $C$. 
The value of the latter is determined by the magnitude
of the asymmetry between bosons and anti-bosons. Notice that in equilibrium
chemical potentials of particles and antiparticles are opposite,
$\bar\mu = -\mu$. Hence if bosons condense with $\mu_B = m_B$, the anti-bosons cannot
condense because $\mu_{\bar B} = -m_B$.

We assume above that the collision amplitude is T-invariant but even if 
this restriction 
is lifted the collision integral is still annihilated by 
functions (\ref{Boson_distr}) or (\ref{Bose_cond_distr})
due to S-matrix unitarity~\cite{ad-cyclic}.

As is stated in the Introduction, a large lepton asymmetry is necessary to make
$W$ bosons condense. 
In particular, the condensate is formed at 
lepton number densities higher than the critical one $n^c_\nu = m_W^3/6 \pi^2$. 
When $T \rightarrow 0$, the $W$-boson mass, $m_W$, approaches its
usual vacuum value, created by the Higgs condensate. When the
temperature rises, there are two opposite effects on $m_W$. The
first one is the usual positive temperature correction 
$\delta m_W \sim e T$. The second effect is negative and is related to a
decrease of the amplitude of the Higgs condensate.  As a result, at
temperatures above the 
electroweak phase transition, when the Higgs condensate disappears \cite{kirzhnits}, 
the sole contribution to $W$-mass comes from the temperature
corrections ~\cite{mW-of-T} and $m_W$ may be 
much smaller than its vacuum value.    

The temperature dependence of $n^c_\nu$ below the EW phase transition 
was analyzed in refs.~\cite{Ferrer:1987jc,Kapusta:1990qc}. 
It should be noted here that, if T is higher than the
critical temperature of the EW phase transition, 
the $W$ condensation may take place also for lower,
but still large, values of the lepton asymmetry~\cite{Kapusta:1990qc}. 
In this case, as is mentioned above, the $W$ mass would 
be essentially determined by the medium effects, i.e. by the
temperature corrections or by the condensate itself. 
Thus the $W$ mass in the electroweak symmetric phase
might be lower than in the broken phase and a lower lepton 
asymmetry would be required for the condensation. 
In this paper we mostly assume that the temperature is below the
electroweak phase transition and thus the plasma consists of massive $W$ and $Z$
bosons, neutral Higgs bosons, quarks, leptons, and their antiparticles.
Nevertheless the interesting possibility of $W$ condensation at lower lepton 
asymmetry should be kept in mind.

Models of generation of large lepton asymmetry were considered in 
refs.~\cite{large-L}-\cite{ad-nu-rev}. 
However, if the asymmetry was generated above
the electroweak (EW) phase transition, it might be transformed into the baryonic one 
by the sphaleron processes, creating unacceptably large baryon asymmetry
So we need to assume that the lepton asymmetry should
be created below the EW phase transition. 

On the other hand, a mechanism to avoid generation of too large baryon 
asymmetry could be triggered by a large lepton asymmetry itself. It was
pointed out in ref.~\cite{linde} and confirmed in several
subsequent papers~\cite{EW-suppress,casas} that a large lepton
asymmetry suppresses restoration of the electroweak symmetry and hence
if lepton asymmetry was  
generated at very high temperatures when sphalerons were not active and
the electroweak symmetry was broken at these high temperatures, the
electroweak baryon non-conservation would never be efficient.

Finally, concerning the formation of the Bose-Einstein condensate, it is essential that 
the particles in question possess a conserved quantum number that forces their 
chemical potential to be 
nonvanishing, if the number of particles is not equal to the number of antiparticles.
The amplitude $C$ is then calculated from the known difference of the number 
densities of particles and antiparticles.

In this paper we consider a simple example of electrically neutral plasma in
which charged $W$-bosons condense because of a large asymmetry between leptons
and antileptons. 
For simplicity we confine ourselves to only one
family of leptons. This simplification does not influence the 
essential features of the result. A more detailed analysis with 
all the particles included
can be found in ref.~\cite{harvey}. Quarks may be essential
for the imposing of the condition of vanishing of all gauge charge densities 
in plasma and for the related cancellation of the axial anomaly but we work 
in the lowest order of the perturbation theory where the anomaly is absent.

The plasma is supposed to be electrically neutral, with zero baryonic number
density but with a high leptonic one. The essential reactions are the direct and 
inverse decays of $W$: 
\begin{eqnarray}
W^+ \leftrightarrow e^+ + \nu\,. 
\label{W-e-nu}
	\end{eqnarray}
The equilibrium with
respect to these processes imposes the equality between the chemical potentials:
\begin{eqnarray}
\mu_W = \mu_\nu - \mu_e
\label{mu-W}
	\end{eqnarray}
The condition of electroneutrality reads:
\begin{eqnarray}
n_{W^+} - n_{W^-} - n_{e^-} + n_{e{^+}} = 0
\label{el-neutral}
	\end{eqnarray}
The leptonic number density is equal to
\begin{eqnarray}
n_L = n_\nu - n_{\bar \nu} + n_{e^-} - n_{e^+}\,.
\label{n-L}
	\end{eqnarray}
Here $n =g_s\int d^3p /(2\pi)^3 f $ is the number density of the corresponding 
particles and $g_s$ is the spin counting factor. One should remember that only
left-handed electrons participate in reaction (\ref{W-e-nu}), chirality is conserved in
reactions with $Z$-bosons and photons, and chirality flip may take place only
in reactions with Higgs bosons.

Leptonic number is supposed to be conserved, so $n_L$ is constant in the comoving
volume and is a fixed parameter of the scenario. (B+L)-nonconservation induced
by sphalerons is neglected because temperature is mostly assumed to be below the electroweak
phase transition.

It is evident that for sufficiently high $n_L$ the chemical potential of $W$
should reach its maximum value, $\mu_W = m_W$, and with further rising $n_L$, 
W-bosons must condense. 

Let us first assume that the density of leptonic charge is very large, $n_{L} > T^{3}$. Correspondingly
the amplitude of the condensate must also be large, $n_{W} ^{C}\approx C/(2\pi)^{3 }> T^{3} $. In this case
$\mu_{e,\nu}> m_{W}$ and equations (\ref{mu-W}--\ref{n-L}) can be easily solved: $\mu_{\nu} \approx \mu_{e}$,
$n_{L} \approx \mu_{\nu}^{3}/2\pi^{2}$, and $n_{W}^{C} \approx (2/3)
n_{L}$. Interparticle separation of $W$-bosons under these conditions
is $d\sim  n_L^{-1/3} < T^{-1}$, while the de Broglie wave length of the
condensed $W$'s with zero momentum is formally infinitely large. In the realistic condensate
the particle momentum is not precisely zero but is of the order of the inverse size, $d^{-1}$, 
of the region where $W$-bosons condense. 
Another relevant 
quantity is the de Broglie wave length of $W$-bosons above the condensate, 
$\lambda_{dB} \sim 2\pi/\sqrt{2m T}$. The condition $\lambda_{dB} > d$ would be
fulfilled as long as $T< m_W/2\pi^2$.

A huge cosmological lepton asymmetry, $n_L \gg T^3$, could be created in the
Affleck-Dine scenario~\cite{affl-dine}.  In this model the universe
could be quite cold. Later when $n_L \sim 1/a^3$ is diluted by the
cosmological expansion down to the value when $\mu_W$ becomes smaller
than $m_W$ the condensate would evaporate and the universe would be reheated.

Though the possibility of 
a huge lepton asymmetry is quite interesting, the condensation
of $W$-bosons could take place even with much smaller 
$n_L \sim m_W^3$, as can be seen from the analysis of the kinetics
presented in this section. 
The interparticle separation of W-bosons under these conditions is 
$d\sim  n_L^{-1/3} \sim m_W^{-1}$.
In this case the temperature may be relatively high, 
$T\simeq m_W$,  but nevertheless $W$-bosons could condense. 
According to the equilibrium distribution (\ref{Bose_cond_distr}) the
plasma would consist of two parts, the condensate with zero momentum
and the high temperature plasma over the condensate. 
The de Broglie wavelength of the high temperature plasma would be again
$\lambda_{dB} \sim 2\pi/\sqrt{2m T}$, that is larger than $d$ as long as $T < 2\pi^2 m_W.$
As is argued
above, the de Broglie wave length of the condensed W-bosons 
even in this case is much larger than the interparticle distance.

Thus at $T\sim m_W$ the charged weak bosons, $W$, condense if 
all the relevant quantities are close by an order of magnitude to the
$W$-boson mass to  a proper power, i.e. $\mu_\nu \sim \mu_e \sim m_W$,
$n_L \sim m_W^3$. They rise with rising temperature, as can be found from
numerical solution of eqs.~(\ref{mu-W}--\ref{n-L}).

We will also discuss the $W$ condensation above the EW phase
transition. In this situation the Higgs condensate is absent and the 
$W$-bosons acquire, due to temperature corrections, 
$m_W \sim g T$~\cite{mW-of-T}, where $g\sim 0.1$ is the typical
coupling constant.

The critical values of chemical potential necessary to achieve $W$ condensation at lower $n_L$ have been calculated, even though an analytic solution is possible only in some special limits and the approximations used are not valid in some regions of parameters. The general equations together with some critical discussion concerning the approximations used can be found in the literature, see \cite{Ferrer:1987jc,Kapusta:1990qc}. For instance, at high temperature, that is $T$ much larger than masses and chemical potentials but smaller than the EW breaking value, 
$W$-bosons would condense if
$	\mu_L > 0.3 \sqrt{T^2 - T_c^2}$,
where $\mu_L$ is the chemical potential associated to the total lepton number and $T_c \sim 200$ GeV \cite{Kapusta:1990qc}. 
Another limit that can be analytically calculated is the low temperature one ($T\sim 0$). Under this hypothesis, 
$W$-bosons condense when $\mu_L > 88.4$ GeV $+ T^2/64.8$ GeV.

A caveat for a large lepton asymmetry arises from the consideration of the
big bang nucleosynthesis with strongly mixed neutrinos. It is shown~\cite{bbn-osc} 
that leptonic chemical potentials of all neutrino flavors are restricted by
$|\mu_\nu /T |< 0.07$ at the BBN epoch. 
However, it should be noted that the entropy release from the electroweak epoch down to the
BBN epoch diminishes the lepton asymmetry, $n_L/n_\gamma$, by the
ratio of the particle species present in the cosmological plasma at
these two epochs, which is approximately $10$. Hence the original lepton 
asymmetry, $(n_L/n_\gamma)_{EW}$ could be  of order unity and still  compatible with BBN.  
It should be also noted that the BBN bound can be avoided~\cite{majoron} 
if neutrinos are coupled to the hypothetical pseudogoldstone boson, Majoron. 

Moreover, at high temperatures, when the Higgs condensate is 
underdeveloped, the $W$-boson mass may be noticeably smaller than the
plasma temperature and $W$ could condense even at $|\mu/T| < 0.07 $.
If $W$s condensed above the electroweak phase transition,
the magnitude of the chemical potential necessary for the condensation
was much smaller than temperature, $\mu_{W} \sim g T$, as explained
above. As we see, in this case the chemical potentials of electrons
and neutrinos would be also much smaller than temperature. In this
limit the differences between number densities of fermions and
antifermions are:
\begin{equation}
n_F - n_{\bar F} = \frac{1}{6}\, g_F \mu_F T^2,
\label{n-bar-n}
\end{equation}
where $g_F$ is the number of the spin degrees of freedom, $g_e =2$,
$g_\nu =1$.

Substituting this expression into eqs.~(\ref{mu-W}--\ref{n-L}) and
assuming an arbitrary chemical potential $\mu_W<T$, we find
that the condensate would be formed if approximately
$n_L \geq g T^3$ 
and that all the chemical potentials are of the order of $gT$. 
Correspondingly the necessary lepton asymmetry could be as small as
$n_L/n_\gamma \sim g$. We see that even
without the entropy release such lepton asymmetry is not dangerous for BBN.

\section{  Equations of motion of gauge bosons and their condensation. 
\label{s-eq-of-mot-W}}

In this section we consider the evolution of the gauge bosons of 
$SU(2) \times U(1)$ electroweak model and analyze the possibility 
of the condensate formation as supplementary to the approach of
the previous section. In addition to that, using the equations of
motion allows to take into account the spin-spin interactions of the
vector bosons and to check if the condensate is ferromagnetic or
antiferromagnetic.

The Lagrangian of the minimal electroweak model  has the well known form:
	\begin{eqnarray}
	\label{SM_tot_L}
	L = L_{gb} + L_{sp} + L_{sc} + L_{Yuk},
	\end{eqnarray}
which are respectively the gauge boson, the spinor, the scalar, and the 
Yukawa contributions. Explicitly we have: 
	\begin{eqnarray}
	\label{SM_kin_L}
	L_{gb} = - \frac{1}{4} G^i_{\mu\nu} G^{i \, \mu\nu} -
	\frac{1}{4} f_{\mu\nu} f^{\, \mu\nu} ,
	\end{eqnarray}
	\begin{eqnarray*}
	G^i_{\mu\nu} = \partial_\mu A^i_\nu - \partial_\nu A^i_\mu + 
	g \epsilon^{ijk} A^j_\mu A^k_\nu,
	\hspace{1.5cm}
	f_{\mu\nu} = \partial_\mu B_\nu - \partial_\nu B_\mu,
	\hspace{1.5cm}
	i=1,2,3
	\end{eqnarray*}
	\begin{eqnarray*}
	\label{L-sp}
	L_{sp} = \bar \Psi i D \!\!\!\!/  \; \Psi,
	\hspace{1cm}
  	D_\mu \Psi = 
  	\left( \partial_\mu - \frac{i}{2} g \, \sigma^j  A^j_\mu - 
        \frac{i}{2} g' Y B_\mu
   	\right) \Psi,
	\end{eqnarray*}
	\begin{eqnarray*}
	L_{sc} = \frac{1}{2} 
	\left( D_\mu \Phi \right)^\dag \left(D_\mu \Phi \right) 
	+ \frac{1}{2} \mu^2 \Phi^\dag \Phi - \frac{1}{4} \lambda (\Phi^\dag \Phi)^2,
	\hspace{1cm}
	D_\mu \Phi = \left( \partial_\mu - \frac{i}{2} g \, \sigma^j A^j_\mu - 
	\frac{i}{2} g' B_\mu
	\right) \Phi,
	\end{eqnarray*}
and the Yukawa Lagrangian describes interactions of fermions with the Higgs 
field.

In the expressions above $A^i_\mu$ and $B_\mu$ are the gauge boson potentials 
of the $SU(2)$ and $U(1)$ groups respectively, $g$ and $g'$ are their
gauge coupling constants, $Y$ is the hypercharge operator
corresponding to the $U(1)$ group and $\sigma^j$ are the Pauli matrices
operating in $SU(2)$ space. As usually, the  repeated indices imply summation.

In the broken phase the physical massive gauge boson fields are obtained
through the Weinberg rotation:
	\begin{eqnarray}
	W^{\pm}_\mu = \frac{A^1_\mu \mp i A^2_\mu}{\sqrt{2}}\, ,
	\hspace{1cm}
	Z_\mu = c_W A^3_\mu - s _W B_\mu\,,
	\hspace{1cm}
	A_\mu = s_W A^3_\mu + c_W B_\mu,
	\end{eqnarray}
where $c_W$ and $s_W$ stand respectively for $\cos \theta_W$ and $\sin\theta_W$ 
and $\theta_W$ is the Weinberg angle.
The other fields involved in the Lagrangians presented above are 
the spinor $\Psi$ and the Higgs field 
$\Phi = [\phi^+, \phi^0]^T$. The latter,
after the symmetry breaking, acquires non-zero vacuum expectation
value, $v$, and takes the form  
$\Phi = (1/\sqrt{2}) [0,v+\phi_1^0 ]^T$, where the upper index $T$ means
transposition and $\phi_1^0$ describes excitations in the broken symmetry
phase, i.e. neutral massive Higgs particle.
We are considering here one lepton family but the results 
can be easily generalized to the three family model.
We use the unitary gauge in which the particle
content is explicit, for example physical gauge bosons have three polarization 
states and only one physical neutral Higgs field, $\phi_1$ is present.

From the presented above equations one can conclude that 
in addition to the usual kinetic and mass terms, vector boson interactions
contain the following
cubic and quartic couplings, see e.g. ref.~\cite{pich}:
\begin{eqnarray}\label{eq:cubic}
L_3 &\!\!\!\! = &\!\!\!\!
i e \cot{\theta_W}\left[
\left(\partial^\mu W^\nu -\partial^\nu W^\mu\right)
 W^\dagger_\mu Z_\nu -
\left(\partial^\mu W^{\nu\dagger} -\partial^\nu W^{\mu\dagger}\right)
 W_\mu Z_\nu +
W_\mu W^\dagger_\nu\left(\partial^\mu Z^\nu -\partial^\nu Z^\mu\right)
\right]
\nonumber
\\[10pt] &&\!\!\!\!\mbox{}
+i e \left[
\left(\partial^\mu W^\nu -\partial^\nu W^\mu\right)
 W^\dagger_\mu A_\nu -
\left(\partial^\mu W^{\nu\dagger} -\partial^\nu W^{\mu\dagger}\right)
 W_\mu A_\nu +
W_\mu W^\dagger_\nu\left(\partial^\mu A^\nu -\partial^\nu A^\mu\right)
\right]\,,
%
\\[8pt] \nonumber 
L_4 &\!\!\!\! = &\!\!\!\!\mbox{}
-{e^2\over 2\sin^2{\theta_W}}\left[
\left(W^\dagger_\mu W^\mu\right)^2 - W^\dagger_\mu W^{\mu\dagger}
W_\nu W^\nu \right]
- e^2 \cot^2{\theta_W}\,\left[
W_\mu^\dagger W^\mu Z_\nu Z^\nu - W^\dagger_\mu Z^\mu W_\nu Z^\nu
\right]
\nonumber
\\[10pt] &\!\!\!\! &\!\!\!\!\mbox{}
- e^2 \cot{\theta_W}\left[
2 W_\mu^\dagger W^\mu Z_\nu A^\nu - W^\dagger_\mu Z^\mu W_\nu A^\nu
- W^\dagger_\mu A^\mu W_\nu Z^\nu
\right]
\nonumber
\\[10pt] &\!\!\!\! &\!\!\!\!\mbox{}
- e^2\,\left[
W_\mu^\dagger W^\mu A_\nu A^\nu - W^\dagger_\mu A^\mu W_\nu A^\nu
\right] .
\label{quartic}
\end{eqnarray}

Using the standard Euler-Lagrange procedure we can obtain the following
Maxwell equations for the electromagnetic field:
\begin{eqnarray}
\partial_\mu F^{\mu}{_\nu} &=& J_\nu^\psi +
ie\left[  W^\dagger_\mu\partial_\nu W^\mu    - W^\mu\,\partial_\nu W^\dagger_\mu
+ W^\dagger_\nu\,\partial_\mu W^\mu -  W_\nu\, \partial_\mu W^{\mu \dagger} 
+2 W^\mu \partial_\mu W^\dagger_\nu - 2 W^{\mu \dagger} \partial_\mu W_\nu \right]
\nonumber \\
&+& e^2 \left[ 2A_\nu W_\mu^\dagger W^\mu - A^\mu \left( W^\dagger_\mu W_\nu + 
W_\mu W^\dagger_\nu\right)
\right] \, \nonumber\\
&+& e^2\cot{\theta_W} 
\left[ 2Z_\nu W_\mu^\dagger W^\mu - Z^\mu \left( W^\dagger_\mu W_\nu + 
W_\mu W^\dagger_\nu\right) \right] \,,
\label{d-F-mu-nu} 
\end{eqnarray}
where $A_\mu$ is the electromagnetic potential, 
$F_{\mu\nu} =\partial_\mu A_\nu - \partial_\nu A_\mu$ is the Maxwell tensor,
and $J^\psi_\nu$ is the electromagnetic current of the charged fermions.
All the rest in the r.h.s. of eq. (\ref{d-F-mu-nu}) can
be understood as the
electromagnetic current of  $W$ bosons, $J_\mu^W$.
In what follows we assume that the total
electric charge density of the plasma is zero and the average
three-vector current vanishes as well.

Equations of motion for the other vector fields , $W^\pm$ and $Z$, can be obtained
from the Lagrangians (\ref{eq:cubic}, \ref{quartic}) plus the contributions from
the kinetic and mass terms:
\begin{eqnarray}
\partial_\mu W^\mu_\nu + m^2_W W_\nu &=& ie \left[ A^\mu W_{\mu\nu} -
\partial_\mu (W^\mu A_\nu) + \partial_\mu (W_\nu A^\mu ) - W^\mu F_{\mu\nu}\right]
\nonumber \\
 &+& ie \cot \theta_W \left[ Z^\mu W_{\mu\nu} + \partial_\mu (Z^\mu W_\nu) -
\partial_\mu (W^\mu Z_\nu) + W^\nu Z_{\nu\mu} \right] \nonumber\\
&+& \left({e^2}/{\sin^2 \theta_W}\right) \left[ W_\nu (W^\dagger_\mu W_\mu)
-W^\dagger_\nu (W^\mu W_\mu) \right] +
e^2 \cot^2 \theta_W \left( W_\nu Z_\mu Z^\mu - Z_\nu  Z_\mu W^\mu  \right)
\nonumber\\
&+& e^2\cot \theta_W \left( 2 W_\nu Z_\mu A^\mu - Z_\nu W_\mu A^\mu - 
A_\nu W_\mu Z_\mu \right) +
e^2 (W_\nu A^\mu A_\mu - A_\nu W_\mu A_\mu ),
\label{dW} \\
\partial_\mu Z^\mu_\nu + m^2_Z Z_\nu &=& 
  ie \cot \theta_W \left[ W^{\mu} W^\dagger_{\mu\nu} - W^{\mu^\dagger} W_{\mu\nu}
+ \partial_\mu (W^\mu W^\dagger_\nu) - \partial_\mu (W^{\mu\dagger} W_\nu)\right]  
\nonumber\\
&+&e^2 \cot^2 \theta_W \left( 2Z_\nu W^\dagger_\mu W^\mu - W^\dagger_\nu  W_\mu W^\mu  
- W_\nu W^\dagger_\mu Z^\mu \right)
\nonumber \\
&+& e^2\cot \theta_W \left( 2 A_\nu W^\dagger_\mu W^\mu - W^\dagger_\nu W_\mu A^\mu - 
W_\nu W^\dagger_\mu A_\mu \right),
\label{dZ}
\end{eqnarray}
where $V_{\mu\nu} = \partial_\mu V_\nu - \partial_\nu V_\mu$, $V_\mu=W_\mu,Z_\mu$ 
and we assumed that the fermionic currents coupled to $W$ and $Z$ are
absent.  To avoid confusion it is worth noting that the field strength defined
after eq. (\ref{SM_kin_L}) includes the gauge boson self-coupling, while it 
is not included into $V_{\mu\nu}$.

Equations (\ref{dW},\ref{dZ}) are the equations for the corresponding field operators.
They describe classical fields in the tree approximation and do not include
the effects of $W$ and $Z$ instability. The latter can be taken into account by
perturbative iteration of these equations. The effects of instability are discussed 
below.

We will consider the case when the electric charge density of fermions, $J_0^\psi$, 
is non-vanishing and homogeneous.
It is usually assumed that the primeval plasma
is electrically neutral and thus the non-zero charge density of fermions must be
compensated by the opposite sign charge density of $W$. To study such
a system  it is convenient to use the electromagnetic gauge freedom and to
impose the condition $A_\mu =0$. We also assume that the average value
of $Z_\mu =0$. In this case there exists a homogeneous solution of the
equation of motion of the form:
\begin{eqnarray}
W_\mu = C_\mu \exp (-i m_W t ),
\label{W-mu}
\end{eqnarray}
where we impose the condition $\partial_\mu W^\mu = 0$ to eliminate the non-physical
spin degrees of freedom. So $C_\mu$ is a constant vector with vanishing time component 
and $m_W$ is the boson mass determined by the nonzero vacuum expectation
value of the Higgs field~\footnote{We want to stress an important 
difference between Bose-Einstein condensate and Higgs condensate,
which is often overlooked. The equation of state of the former is simply
$P=0$, while for the latter $P=-\rho$, where $P$ and $\rho$ are respectively
pressure and energy densities.}. 
In contrast,  in some papers the gauge
condition of time independent charged vector field
is taken: $W_\mu = C_\mu$, 
while the electromagnetic
vector potential is non-zero: $A_\mu = \delta_\mu^t m_W/e$.

In addition to the Higgs induced part, the mass of $W_\mu$ contains
also the contributions induced by the temperature
effects~\cite{mW-of-T} and by the impact of the $W$-condensate
itself, which are disregarded at this stage, see discussion in
sec.~\ref{s-kinetics} about possible condensation of $W$ at lower
lepton asymmetry.

Solution (\ref{W-mu}) corresponds to the Bose condensate of $W$-bosons
describing a collection of these 
particles at rest. This field configuration corresponds to the condensation
of the positively charged $W^+$. The condensation of $W^-$ is
described by the complex conjugate expression.
However, such a solution is not obligatory and depends 
upon the kinetics of the system and the interactions between $W$-bosons
at rest. In fact the only condition which we have at this level is the
condition of the electric neutrality and it demands only that the average values
of bilinear combinations of $W$ must be non-zero, while classical vector
field $W_\mu$ may vanish on the average. A possible vanishing of the classical
vector field $W_j$, where $j= 1,2,3$ is the spatial vector index, is
physically quite evident. The condensate is a collection of $W$-bosons
at rest with the charge density which compensates the charge
density of fermions. The most probable state of such particles (the
highest entropy state), if the spin-spin interaction is neglected, is
the state with chaotic distribution of the individual spins. It is
natural to expect that the average value of the total spin in such a
state is zero or at least not macroscopically large. The situation is
opposite in the ferromagnetic case when the spin alignment is energetically
favorable and the classical vector field could be formed.

In reference~\cite{linde} a similar statement of the formation of
a classical vector field was done but without an analysis of the
dynamics introduced by the spin-spin interaction. In the quoted paper the 
mentioned above  gauge of time independent $W$ is used,
but the physical results are of course gauge invariant. In this gauge
the condition of the charge neutrality becomes
$2 A_0 W^+_\mu W^-_\mu = - J_0^\psi$, which is again bilinear in $W$
field and from the condition of non-zero charge density of the
condensed $W$'s does not follow that there exists
the classical field  $W_j \neq 0$. 

One more comment may be in order here. The instability of $W$ can be
introduced in the usual way by adding an imaginary  
part to the mass equal to the decay half-width. The introduction of such
a term into the equation of motion for $W$ leads to the exponential decay
of the field, $W\sim \exp(-\Gamma t/2)$. However this description is valid
only for the decay into vacuum. 

For the decay into a dense medium the Fermi
exclusion principle should be taken into account. Hence, if neutrinos have
sufficiently large chemical potential, such that
all the states where $W$ could decay would be
occupied, the decay rate would be exponentially suppressed and solution (\ref{W-mu})
could be physically realized. This observation establishes the equivalence
between the kinetic approach of sec. \ref{s-kinetics} and that presented here.

In fact the absolute stability of the condensed $W$'s is
unnecessary.  Even if $W$-bosons decay, 
the ferromagnetic condensate
can be formed, if the time of the
condensate formation is shorter than the life-time of $W$-bosons in
the plasma. The condensed $W$-bosons are in the state of dynamical
equilibrium: $W$'s evaporating from the condensate because of their decay
or scattering of hot fermions, are compensated by $W$'s coming
back by the inverse processes. In thermodynamical equilibrium the
average number of $W$-bosons in the zero momentum state remains
constant. The decay rate of the $W$-bosons in plasma is proportional
to
\begin{equation}
\Gamma \sim \frac{\Gamma_0} {\left[ e^{-(m_W/2 + \mu_e)/T} +1\right]
\left[ e^{-(m_W/2 - \mu_\nu)/T} +1\right]}\,,
\label{Gamma-W}
\end{equation}
where $\Gamma_0 \sim \alpha m_W$ is the decay rate of $W$ in vacuum.
The 
denominator in eq. (\ref{Gamma-W}) comes from the Fermi
suppression terms $(1-f_{e^+})(1-f_\nu)$. 
We consider the decay 
$W^+\rightarrow \nu e^+$ and take into account that in thermal
equilibrium the chemical potentials of electrons and positrons are
equal by magnitude but have the opposite signs. In the case of very
large lepton asymmetry, $n_L$, $\mu_\nu \gg T$, the decay rate of
$W$-bosons would be exponentially
suppressed and their life time can be longer than the time of the spin
alignment, $\tau_S$. 
As we mentioned above,
the cosmological generation of $L\gg T^3$ may be
realized in a version of the Affleck-Dine scenario~\cite{affl-dine}
which leads to a cold universe with non-relativistic $W$-bosons.
Moreover, even in absence of the exponential suppression and
relatively small
lepton asymmetry, $L\sim T^3$, the life-time of $W$-bosons in
the plasma can be larger 
than $\tau_S$. As is shown below, the Hamiltonian
of spin-spin interaction is given by
eqs. (\ref{U_spin},\ref{U-spin-W}).  
Correspondingly the characteristic time of the spin alignment can be
estimated as:
\begin{equation}
\tau_S \sim 1/U^{spin} \sim m_W^2/( n_W e^2 S^2)\,,
\label{tau-S}
\end{equation}
where $S$ is the total spin of the condensed $W$-bosons, $n_W \equiv 1/d^3$ 
is their number density, and $d$ is the average distance between them.
We approximated $\delta (\bf r)$ as $1/d^3 = n_W$.
Evidently $\tau_S$ can be considerably shorter than the life-time of
$W$-bosons.
The same ``stability'' arguments are applicable to the decay of $W$
into quarks.

\section{Spin-spin interactions of $W$ bosons \label{s-spin}}

As we mentioned above, the form of the vector boson condensate depends upon
the interaction between the vector bosons at rest. 
If the latter favors the opposite spin configuration,
i.e. a pair of bosons ``prefers'' to be in the zero spin state,
the condensate would have zero total spin, i.e. $W$-bosons would form
the scalar condensate (antiferromagnetic case). In the opposite
situation of the favorable spin-two state, the spins of all vector
bosons in the condensate would be 
aligned and the condensate 
would have macroscopically large spin (ferromagnetic case). 

First, in subsection \ref{ss-elmagn} we will consider 
only the spin-spin interaction induced by the electromagnetic interactions of
$W$-bosons, namely by the coupling of their magnetic moments. Next, in 
subsection \ref{ss-W4} we will take into account the local quartic 
self-coupling of $W$. In subsection \ref{ss-WZ} we take into account
the spin-spin interaction caused by the $Z$-boson exchange between $W$'s.
In sec. \ref{ss-screen} we discuss the effects of possible screening of
the magnetic moment interactions by plasma effects. In
section~\ref{ss-discuss} we discuss the results of the previous
subsections and conclude.

\subsection{Electromagnetic interactions \label{ss-elmagn}}

The essential particles in the system we study in this section are the
charged $W$ bosons at rest and charged relativistic fermions,
which ensure the electric neutrality of the medium. The latter are
electrons (or positrons) and quarks but these details are not important.
For relativistic fermions helicity is conserved and 
hence the interaction of their spins with the spins of $W$ is not essential, because
on the average the electron-positron medium is not spin-polarized.  
Accordingly  we take into account only the spin-spin interaction of
non-relativistic $W$-bosons and disregard the impact of the charged
fermions. The electromagnetic interaction between $W$-bosons is 
similar to the well known 
interaction of nonrelativistic electrons, which is described by
the Breit equation. The derivation of this equation can be found e.g. in
book~\cite{Landau4}. According to the Breit equation the interaction between
the magnetic moments of two electrons (i.e. their spin-spin interaction) 
has the form:
	\begin{eqnarray}
	\label{U_M_Breit}
	U^M(r) = \frac{e^2}{16\pi m_e^2}
	\left[ 
	\frac{(\mathbf{ \sigma_1} \cdot \mathbf{ \sigma_2})}{r^3}
	- \frac{3 (\mathbf{ \sigma_1} \cdot \mathbf{r}) (\mathbf{ \sigma_2} \cdot 
\mathbf{r}) }{r^5}
	- \frac{8\pi}{3} ( \mathbf{ \sigma_1} \cdot \mathbf{ \sigma_2}) \delta^{(3)} 
(\mathbf{r})
	\right]\,,
	\end{eqnarray}
where $\sigma_{1,2}$ are the spin operators of the electrons, i.e. the
Pauli matrices averaged respectively over the first and second
electron wave functions, and $e^2 = 4\pi \alpha = 4\pi/137 $ (to avoid
confusion let us note
that in ref.~\cite{Landau4} the notation is different, namely $e^2 = \alpha$).

The expression for potential (\ref{U_M_Breit}) created by the one photon exchange
is true for virtual photons propagating in vacuum. The presence of
plasma changes the propagator and could modify the spin-spin
potential. This is discussed in section~\ref{ss-screen}.  The
screening effects are not important, at least in some
temperature range. 

The analogue of the Breit equation for $W$-bosons can be derived along
exactly the same lines as is done for  electrons. The electromagnetic
interaction between two $W$ bosons in the lowest order in the electric
charge, $e$,  is described by the usual one-photon exchange diagram.
In the Feynman gauge, where the photon propagator is 
$D^{\mu\nu} =  g^{\mu\nu}/ q^2$, the amplitude corresponding to
this diagram is:
	\begin{eqnarray}
	\label{V_M_ampl_i}
	M = - \frac{1}{(p_1 - p_2)^2}  W^{\alpha' \dagger} W^{\beta' \dagger}
	V_{\alpha' \alpha \mu} (p_1, p_2,q) V_{\beta' \beta}\;^{\mu} (p_3, p_4,-q)
	W^\alpha W^\beta 
	\end{eqnarray}
where $V_{\alpha \beta \mu}$ is the most general CP invariant
$W^\dagger W\gamma$ vertex \cite{Bardeen:1972vi}:
	\begin{eqnarray}
	\label{V_3}
	V_{\alpha \beta \mu} (p_1, p_2,q) =
	ie \left[ p_\mu g_{\alpha \beta} 
	+ 2 \left( q_\beta g_{\alpha \mu} - q_\alpha g_{\beta \mu} \right)
	+ (1-k_\gamma) \left( q_\beta g_{\alpha \mu} - q_\alpha g_{\beta \mu} \right) 
	+ \left( \frac{\lambda_\gamma}{2m_W^2} \right) p_\mu q_\alpha q_\beta 
	\right],
	\end{eqnarray}
where $p_1$ and $p_3$ are the momenta of the incoming particles, 
$p_2$ and $p_4$ are the momenta of the outgoing particles and $p = p_1
+ p_2$, $q = p_2 - p_1$. This expression should be 
symmetrized with respect to the interchange of $W$-bosons in the
initial and/or in the final states. 

The vertex written above contains
two anomalous coupling parameters $k_\gamma$ and $\lambda_\gamma$.
As we can see from eq.  (\ref{d-F-mu-nu}),
the standard electroweak model predicts, up to the second order in the 
electromagnetic coupling constant 
$e$: $k_\gamma =1 $, $\lambda_\gamma = 0$. 
In what follows we mostly assume that these values are true, since
they are compatible  with the present experimental data for triple
gauge boson couplings \cite{Schael:2004tq}. In this case the amplitude  
(\ref{V_M_ampl_i}) reduces to:
	\begin{eqnarray}
	\label{V_M_ampl_e}
	M = \frac{e^2}{q^2} 	W_{1'}^{\alpha' \dagger} W_{2'}^{\beta' \dagger}
	\left[ p_\mu g_{\alpha \alpha'} + 2 (q_\alpha g_{\alpha' \mu} - q_{\alpha'} g_{\alpha \mu}) \right]
	\left[ p^\mu g_{\beta \beta'} - 2 (q_{\beta} g_{\beta'}\,^\mu - q_{\beta'}   g_{\beta}\,^\mu ) \right]
	W_1^{\alpha} W_2^{\beta } .
	\end{eqnarray} 

The spin-spin interaction is contained in the product of the last two
terms in the square brackets in eq.  (\ref{V_M_ampl_e})  i.e. the terms
containing vector $q$.  The spin operator of vector particles is
defined as the generator of the rotation
group belonging to its adjoint representation and is equal to the
vector product:
	\begin{eqnarray}
   {\bf S}_1 = - i \,\,{\bf W}^\dagger_{1'} \times {\bf W}_1
\label{S}	
\end{eqnarray}
Hence the scattering amplitude induced by the interaction between the
magnetic moments of the charged vector bosons is equal to:
	\begin{eqnarray}
	\label{M_S_mom}
	M_{S } = - \frac{e^2\rho^2}{m_W^2 q^2}
	\left[ 
	q^2 \left(\mathbf{S}_1 \cdot \mathbf{S}_2 \right)
	- \left(\mathbf{S}_1 \cdot \mathbf{q} \right)  \left(\mathbf{S}_2 
\cdot \mathbf{q} \right)
	\right]
	\end{eqnarray}
where $\rho$ is the ratio of the real magnetic moment of $W$ to its
value predicted by the standard electroweak theory $(e^2/m_W^2)$ and
we divided by $4m_W^2$ for proper
normalization of the $W$-wave function, as is explained below, see Sec. \ref{ss-W4}.

The potential which describes the electromagnetic spin-spin
interaction is the  Fourier transform of amplitude (\ref{M_S_mom})
and is equal to:
	\begin{eqnarray}
	\label{U_spin}
	U^{spin}_{em} (r)=  \frac{e^2 \rho^2}{4\pi m_W^2} 
	\left[
	\frac{\left(\mathbf{S}_1 \cdot \mathbf{S}_2 \right)}{r^3}
	- 3 \, 
	\frac{\left(\mathbf{S}_1 \cdot \mathbf{r} \right)  
\left(\mathbf{S}_2
 \cdot \mathbf{r} \right)}{r^5}
	- \frac{8 \pi}{3} \left(\mathbf{S}_1 \cdot \mathbf{S}_2 \right) 
\delta^{(3)}(\mathbf{r})
	\right].
	\end{eqnarray}
This potential has the same form as the corresponding one in the Breit's equation for 
electrons (\ref{U_M_Breit}) but with the different numerical coefficient. 

To calculate the contribution of this potential into the energy of two $W$-bosons
at rest we have to average it over their wave function. In particular, in the condensate case,
it is a S-wave function that is angle independent. Hence the
contributions of the first two terms in eq.~(\ref{U_spin})  
mutually cancel out and only the third one remains, which has negative
coefficient. Thus the energy shift induced by the spin-spin
interaction is equal to:
	\begin{eqnarray}
	\label{En_shift}
	\delta E = \int \frac{d^3r}{V}  \, U^{spin}_{em} (r) = 
	- \frac{2 \, e^2 \rho^2}{3 \, V m_W^2} \left(\mathbf{S}_1 \cdot \mathbf{S}_2 \right),
	\end{eqnarray}
where V is the normalization volume.

Since $S_{tot}^2 = (S_1+S_2)^2 = 4 + 2 S_1 S_2$, the average value of $S_1S_2$ is
equal to
	\begin{eqnarray}
	\label{S-tot}
	S_1 S_2 = S_{tot}^2/2 - 2\,.
	\end{eqnarray}
For $S_{tot} = 2$ this term is $S_1 S_2 = 1 >0$, while for $S_{tot} =0$ it
is $S_1 S_2 = -2 <0$. Thus, if the spin-spin interaction is dominated
by the  interactions between the magnetic moments of $W$ bosons,
the state with their maximum total spin
is more favorable energetically and $W$-bosons should condense in the
ferromagnetic state. This could lead to the spontaneous magnetization in the early 
universe.  

\subsection{Quartic self-coupling of $W$ \label{ss-W4}}

The contribution to the spin-spin interactions of $W$ comes from
the first term in Lagrangian (\ref{quartic}) or from the
third term in the r.h.s. of equation of motion (\ref{dW}). The first term in 
Lagrangian (\ref{quartic}) can be rewritten as:
\begin{eqnarray}
L_{4W}= - \frac{e^2}{2\sin^2\theta_W}\left[(W_\mu^\dagger W^\mu )^2 - 
W_\mu^\dagger W^{\mu \dagger} W_\nu W^\nu \right]=
\frac{e^2}{2\sin^2\theta_W} \left({\bf W}^\dagger \times{\bf W}\right)^2.
\label{spin-spin-W}
\end{eqnarray}
It is assumed here that $\partial_\mu W^\mu =0 $ and thus only the spatial 3-vector
${\bf W}$ is non-vanishing, while $W_t =0$.

Since the corresponding Hamiltonian is obtained from $L_{4W}$ by changing 
sign and since spin operator (\ref{S}) contains imaginary unit factor, the 
sign of the Hamiltonian is positive. It means that the low spin states are energetically
favorable. 

For the comparison of this Hamiltonian with potential energy (\ref{U_spin}) 
we need to properly normalize the wave functions of $W$. In the Hamiltonian
the usual relativistic normalization is used, according to which
the number density of $W$ is equal to $n_W = 2m_W {\bf W^\dagger} {\bf W}$,
while in the non-relativistic Schroedinger equation the wave function is 
normalized to unity,
\begin{eqnarray}
\int d^3 r |\psi|^2 = 1 
\label{nomr-non-rel}
\end{eqnarray}
Accordingly the Hamiltonian should be divided by $4m_W^2$ and its Fourier 
transform producing the spin-spin interaction potential would be:
\begin{eqnarray}
U^{(spin)}_{4W} = \frac{e^2}{8 m_W^2 \sin^2 \theta_W} \left( {\bf S }_1 
{\bf S}_2 \right) \delta^{(3)} ({\bf r}).
\label{U-spin-W}
\end{eqnarray}
Thus the quartic self-coupling of $W$
contributes only to the spin-spin interaction whose sign is
antiferromagnetic.

The same result can be obtained from equation of motion (\ref{dW}) if one 
takes into account that in the non-relativistic limit: 
\begin{eqnarray}
\partial_t^2 {\bf W} = (-E^2 + m_W^2) {\bf W} \approx 2m_W \epsilon {\bf W},
\label{e-non-rel}
\end{eqnarray}
where $\epsilon =( E - m_W)$ is the non-relativistic energy.

\subsection{$Z$-boson exchange \label{ss-WZ}}

The contribution to the spin-spin potential between a pair of $W$ from
the $Z$-boson 
exchange can be found from eq. (\ref{dW}) where we substitute the
expression for $Z_{\nu} $ taken from eq. (\ref{dZ}) in the limit of
vanishing four-momentum of $Z$. Indeed the transferred momentum is
much smaller than $m_Z$, and so the diagram with $Z$-exchange is
effectively local with Z-boson propagator equal to $1/m_Z^2$. 
Hence the contribution from the $Z$-exchange in the $e^2$ order to
eq. (\ref{dW}) is:
\begin{eqnarray}
\partial_\mu W^\mu_i + m_W^2 W_i + 4 e^2 \cot^2 \theta_W 
(m_W / m_Z )^2 ({\bf W^\dagger} {\bf W} ) W_i + ... = 0\,,
\label{Z-to-W}
\end{eqnarray}
where $j = 1,2,3$ is the spatial vector index. 

We see that the $Z$-boson exchange does not contribute to the
spin-spin interactions of $W$. 
However, it should be kept in mind that this result is true only for
the non-relativistic $Z$-bosons, while above the phase transition the
contributions of $Z$ bosons and photons are similar.

\subsection{Plasma screening \label{ss-screen}}

In plasma the time-time component of the photon propagator  
is modified as $1/{\bf q}^2 \rightarrow 1/[{\bf q}^2 + \Pi_{00}(\mathbf{q})]$,
where $\Pi_{00}(\mathbf{q})$ is the photon polarization operator in plasma. Usually
$\Pi_{00} = m_D^2$, where $m_D$ is the Debye screening mass,
which is independent on $\mathbf{q}$. So the
pole at $q^2 =0$ shifts to an imaginary $q$ leading to the well known
effects of the Debye screening. As it was found 
recently~\cite{Dolgov:2008pe} - \cite{Gabadadze:2009jb},
the presence of the charged Bose condensate drastically changes the
polarization operator leading to
an explicit dependence of $\Pi_{00}$ on $\mathbf{q}$
which gives rise to infrared singular terms. The
modification of the propagator takes place already in the lowest order in the
electromagnetic coupling, $e^2$, i.e. in the one loop approximation.

On the other hand, the space-space component of the propagator remains
massless,  $\Pi_{ij} \sim 1/{\bf q}^2$. 
It is known to be true in pure Abelian electrodynamics 
in any order of perturbation theory, while in
non-Abelian theories the screening may occur in higher orders of
perturbation theory due to the infrared singularities~\cite{linde-screen}.
The screening may potentially change the relative strength of the 
electromagnetic spin-spin coupling, which is affected by screening effects, with
respect to local $W^4$ coupling which is not screened. 
However, in the broken phase the system is
reduced to the usual electrodynamics, where screening is absent and
$W$-bosons would condense in the ferromagnetic state. In the unbroken phase
of the electroweak theory the answer is not yet known. Perturbative
calculations are impossible because of the violent infrared
singularities. Maybe lattice calculations analogous to those done in
QCD will help.

The potential describing the magnetic spin-spin interaction is related to
amplitude (\ref{M_S_mom}) with a modified photon propagator. Thus it can be
written as: 
\begin{eqnarray}
U^{(spin)}_{em} ({\bf r})= - \frac{e^2\rho^2}{m_W^2}
\int \frac{d^3 q}{(2\pi)^3} \frac{\exp \left(i {\bf  qr}\right)}{(q^2+ \Pi_{ss}(\mathbf{q}))}
	\left[ 
	q^2 \left(\mathbf{S}_1 \cdot \mathbf{S}_2 \right)
	- \left(\mathbf{S}_1 \cdot \mathbf{q} \right)  \left(\mathbf{S}_2 
\cdot \mathbf{q} \right)
	\right]\,,
\label{U-em}
\end{eqnarray}
where $\Pi_{ss}$ is the plasma correction to the space-space component
of  the photon propagator. 

If, as above, we assume that the wave function of $W$-bosons is space
independent and average this potential over space, we obtain
the following expression for the spin-spin part of the energy shift:
\begin{eqnarray} 
\delta E =\int \frac{d^3 r}{V} U^{(spin)}_{em}  ({\bf r}) =
- \frac{e^2 \rho^2}{V m_W^2} 
\int \frac{d^3 q}{(2\pi)^3} \delta^{(3)} \! ({\bf q}) \,\,
\frac{ q^2 ({\bf S}_1\cdot {\bf S}_2)  - ({\bf  q \cdot S}_1) ({\bf q \cdot S}_2)}
{q^2 + \Pi_{ss} ({\bf q})}
\label{delta-of-q}
\end{eqnarray}
Clearly $\delta E$ vanishes if $\Pi_{ss}$ is non-zero at $q = 0$.
Of course, this is an unphysical conclusion, because the integration
over $r$ should be done with some finite upper limit, $r_{max}=l$, 
presumably equal to the average distance between the $W$ bosons.
So instead of the delta-function, $\delta^{(3)} ({\bf q})$, we obtain:
\begin{eqnarray}
\int_0^l d^3 r \exp (i{\bf qr}) = \frac{4\pi}{q^3}  \left[ \sin (ql) - 
ql \cos (ql) \right].
\label{int-dr}
\end{eqnarray}
The energy shift is given by the expression:
\begin{eqnarray}
\delta E =
- 4 \pi \frac{e^2 \rho^2}{V m_W^2}
S_1^i S_2^j
\int \frac{d^3q}{(2\pi)^3} \,
\frac{\left[\sin(ql) - ql \cos(ql)\right] \left[q^2 \delta_{ij} - q^i q^j \right]}
{q^3\left[ q^2 + \Pi_{ss}({q}) \right]},
\label{delta-E}
\end{eqnarray}
where $V = 4 \pi l^3 /3$.

When we average over an angle independent wave function, e.g. S-wave for the condensate, the non-vanishing part of the integral in Eq. (\ref{delta-E}) is proportional to the Kronecker delta, hence:
	\begin{eqnarray}
	\delta E 
	= S_1^i S_2^j A \, \delta^{ij},
	\end{eqnarray}
where the coefficient $A$ can be calculated by taking trace of Eq. (\ref{delta-E}):
	\begin{eqnarray}
	Tr (A \delta^{ij}) = 3A = 
	- 8 \pi \frac{e^2 \rho^2}{m_W^2} 
	\int \frac{d^3q}{(2\pi)^3} \,
	\frac{\left[q\sin(ql) - q^2l \cos(ql)\right] }
	{q^2\left[ q^2 + \Pi_{ss}({q}) \right]}.
	\end{eqnarray}
Hence the energy shift of a pair of $W$-bosons in S-wave state
due to the spin-spin interaction is:
	\begin{eqnarray}
	\delta E = - \left(\mathbf{S}_1 \cdot \mathbf{S}_2 \right) 
	 \frac{8 \pi e^2 \rho^2}{3 V m_W^2} 
	\int \frac{d^3q}{(2\pi)^3} \,
	\frac{\left[\sin(ql) - ql \cos(ql)\right] }
	{q\left[ q^2 + \Pi_{ss}({q}) \right]},
	\end{eqnarray}
Introducing the new integration variable $x= ql$ we can rewrite it as:
\begin{eqnarray}
\delta E
= - ({\bf S}_1 \cdot {\bf S}_2) \frac{4 e^2 \rho^2}{3\pi V m_W^2}
\int_0^\infty \frac{dx }
{x^2 + l^2\Pi_{ss}(x/l)} \left[x\,\sin x + l^2\Pi_{ss}(x/l) \cos x\right]\,,
\end{eqnarray}
We used here the usual regularization of divergent integrals: 
$\exp(\pm i ql) \rightarrow \exp(\pm i ql - \epsilon q)$ with
$\epsilon \rightarrow 0$. 
With such regularization $\int_0^\infty dx \cos(x) = 0$.

Evidently, if $\Pi_{ss} = 0$, we obtain the same result as that found
in sec.~\ref{ss-elmagn}.  In fact the necessary condition for
obtaining the ``unscreened'' result is $l^2 \Pi_{ss}(x/l) \ll 1$, but
for a large $l^2 \Pi_{ss}$ the electromagnetic 
part of the spin-spin interaction can be suppressed enough to change the
ferromagnetic behaviour into the antiferromagnetic one. This might take place
at high temperatures above the EW phase transition when the Higgs
condensate is destroyed and the masses of $W$ and $Z$ appear only as
a result of temperature and density corrections and thus are
relatively small. The quantitative statement depends upon the
modification of the space-space part of the photon propagator in
presence of the Bose condensate of charged $W$. As far as we know,  this
modification has not yet been found.

\subsection{ Discussion \label{ss-discuss}}

It is instructive to compare the magnetic properties of solid state
bodies with  the considered here (anti-)ferromagnetism of BEC of
$W$-bosons. Magnetic properties of matter are determined 
by the state of outer (unpaired) electrons. A pair of electrons 
belonging to different atoms may be either in
symmetric, $S_{tot} =1$, or antisymmetric, $S_{tot} =0$,
spin state. Since the total wave function of
two electrons must be antisymmetric, their spin state has opposite symmetry with
respect to their orbital wave function. Symmetric and antisymmetric electron states
evidently have different energies, which we denote as $E_s$ and $E_a$ respectively.
Accordingly the spin Hamiltonian can be written as 
	\begin{eqnarray}
	H^{spin} = - J \, \mathbf{S_1} \cdot \mathbf{S_2}.
	\end{eqnarray}
The quantity $J = (E_s - E_a)$ is usually called the exchange energy. Its sign is
determined by the atomic ground state structure.  Evidently
$J>0$ favors parallel spins, while $J<0$ favors antiparallel spins. 
In atomic systems the exchange energy at small distances is typically of the order of 
fractions of eV, that is about $10^3$ times larger than the typical direct 
magnetic dipole (spin-spin) interaction between electrons. 
Hence the exchange interaction
may force the magnetic dipoles of electrons to be aligned or anti-aligned
independently on their direct magnetic interaction. The situation changes at large 
distances, since the exchange energy decays exponentially, while the magnetic 
interaction behaves like $r^{-3}$. Thus the latter dominates on macroscopic scales, 
leading to the breaking of the system
into domains with different directions of the magnetic field
and consequently  to zero net macroscopic 
magnetization. Nevertheless, the ferromagnetic nature emerges when an external
magnetic field is applied to the system.

Fortunately, the system we are considering here is much simpler. All W-bosons
are in the same state with zero momentum and are not binded into a complicated 
atomic system. Evidently the exchange forces are not essential in this case. 
Indeed, $W$-bosons are in symmetric orbital state. Hence their spin wave function
should also be symmetric and both the allowed spin states of $W$,
the scalar, $S=0$, and the tensor, $S=2$, ones are also symmetric.

In principle, electrons and positrons could distort the spin-spin
interactions of $W$  by their spin or orbital motion and
thus destroy the attraction of parallel 
spins of $W$. However, it looks hardly possible because electrons are 
predominantly ultra-relativistic and they cannot be attached to any single $W$ boson
to counterweight its spin. The low energy electrons cannot be long in such a state
because of fast energy exchange with the energetic
electrons. The scattering of electrons (and quarks) on $W$-bosons may lead to
the spin flip of the latter, but in thermal equilibrium
this process does not change the average value of the spin of the condensate.

\section{Conclusion \label{s-conclusion}}
\label{conclusion}

We have studied here the Bose-Einstein condensation of charged weak bosons, $W^\pm$.
Such condensation might occur in the early universe if the
cosmological lepton asymmetry was sufficiently large. The primeval
plasma is supposed to be electrically neutral 
and to have zero density of weak hypercharge. 

In general a Bose-Einstein condensate of vector fields may form either
a scalar state, when the average value of  
vector $W_j$ is microscopically small, or a classical vector state when
all the individual spins of the condensed vector bosons 
are aligned at a  macroscopically large scale. 

In solid state physics the realization of one or the other of these two 
possibilities depends on  the short range exchange forces 
which dominate over the direct spin-spin interaction
at small distances. In the case of $W$ bosons the choice between 
ferromagnetic or antiferromagnetic state is determined by 
the spin-spin interaction of the individual $W$-bosons, realized through
the interactions of their magnetic moments and  their quartic
self-coupling. 

The total spin-spin interaction potential for $W$ is the sum of two terms
(\ref{U_spin}) and (\ref{U-spin-W}). 
As we have discussed below eq. (\ref{U_spin}), the first two terms 
in the interaction of the magnetic moments
cancel each other after averaging over a S-wave function. Thus only
the $\delta $-function term survives.
In the standard electroweak model $\rho = 1$ and thus
the absolute value of the coefficient in front of 
$ {\bf S}_1 {\bf  S}_2 \delta^{(3)} ({\bf r})$ in
eq. (\ref{U_spin}) is $2e^2/(3 m_W^2)$ which is larger than the 
corresponding term in eq. (\ref{U-spin-W}). Hence the former dominates 
and the energetically favored configuration of a multi-W state should have
a macroscopically large total spin. 
However, as we have pointed out in Sec.\ref{ss-screen},
the plasma screening of the interaction between the magnetic moments
may be dangerous for the W-ferromagnetism. 
In the broken phase the problem is reduced to that of pure QED, where
it is known that magnetic forces are not screened. However, in
non-Abelian gauge theories the absence of screening is known only in
the lowest order in perturbation theory, while higher order
calculations suffer from strong infrared divergences and are not
reliable. For the resolution of this problem non-perturbative methods,
as e.g. lattice calculations, are necessary. At the moment the problem
remains unsolved. 

It should be also noted that,
even though Eqs. (\ref{U_M_Breit}) and (\ref{U_spin}) are 
calculated for electrons and $W$ bosons, they are
valid respectively for any spin-(1/2) and spin-1 species having
the usual electromagnetic interactions. 
So they may be applied to other particles, 
including the ones present in the extensions of the standard model. 
If these particles are in a S-wave state, 
on the average the only delta term survives and hence they may form ferromagnetic states.
Indeed, as one can see from Eq. (\ref{S-tot}), the lowest energy state for a S-wave function
is the one with the maximum $S^2_{tot}$.

It is clear from Eqs. (\ref{U_spin}) and (\ref{U-spin-W}) that the
condensate of $W$-bosons would
be antiferromagnetic if the bosons have a negative non-standard
contribution into the magnetic
moment, such that $\rho < 3/(16\sin^2\theta_W)$. 
The ferromagnetism of $W$ can be destroyed also in a model with a smaller
value of the Weinberg angle. All that demands a strong deviation from
the standard model and most probably is excluded, but these effects
may be important in applications to extensions of the
standard model, e.g. SUSY.

If a ferromagnetic state is formed, we would expect that the primeval plasma, where such
bosons  condensed (maybe due to a  large cosmological lepton
asymmetry), can be spontaneously magnetized. The typical size of the magnetic
domains is determined by the cosmological horizon at the moment of the
condensate evaporation. The latter takes place when the neutrino chemical
potential, which scales as temperature in the course of cosmological
cooling down, becomes smaller than the $W$ mass at this temperature.
However, if during the electroweak phase transition a very strong
chaotic magnetic field was generated, the alignment of the spins of
$W$-bosons and the domain size would be determined by this magnetic
field. In the course of the cosmological expansion such field would
drop down as the cosmological scale factor squared
and the spins of $W$-bosons would behave as considered above.
That is, their dynamics would be determined by the
spin-spin interactions and microscopically small
magnetic domains would rearrange themselves into macroscopically large
ones in the same way as it happens in the usual ferromagnets.

Large scale magnetic field created by the ferromagnetism of $W$-bosons
might survive after the decay of the  condensate due to the 
conservation of the magnetic flux in plasma with high electric conductivity.
Such magnetic fields at macroscopically large scales may be the seeds of the
observed larger scale galactic or intergalactic magnetic fields. This
problem will be studied elsewhere, we only note here that
the mechanism of generation of galactic or intergalactic magnetic
fields is unknown and presents a long standing cosmological problem,
for reviews see e.g. ref.~\cite{magn-rev}. Seed magnetic fields generated
during inflation could be quite easily uniform at galactic or
intergalactic scales, but they are too weak, hence a  huge galactic dynamo
is necessary to amplify them up to the observed magnitude. 
However, it is impossible in this way
to explain the existence of intergalactic fields.
On the other hand, seed fields generated at later stages of cosmological
evolution could be quite large (e.g. magnetic fields created at some
cosmological phase transitions), but the characteristic scale of such
fields is by far smaller than the galactic one. The mechanism
suggested here may generate large magnetic fields and
at scales which are of
the order of cosmological horizon at the electroweak temperatures.
In this sense the mechanism is unique. 
Still even after the cosmological stretching
up of the characteristic wave length of the field, it remains smaller than the galactic radius.
Nonetheless, with the ``Brownian motion'' reconnection of the field lines, the
characteristic scale can be enlarged up to the galactic scale, though  by an
expense of their magnitude. Nevertheless with rather mild galactic
dynamo the observed magnetic field may be generated.

{\vspace{0.8cm} \noindent \bf Acknowledgments}
G.P. acknowledges the support provided from the grant program PAPIIT-UNAM IN112308.


\end{document}